\documentclass[aps,prl,twocolumn,groupedaddress,showpacs,superscriptaddress]{revtex4}
\usepackage{graphicx}
\usepackage{wasysym}
\usepackage[dvips]{color}

\begin{document}
\title{Periodic Anderson model with electron-phonon correlated conduction band}

\author{Peng Zhang}
\affiliation{Department of Physics and Astronomy, Louisiana State University, Baton Rouge, Louisiana 70803, USA}
\author{Peter Reis}
\affiliation{Department of Physics and Astronomy, Louisiana State University, Baton Rouge, Louisiana 70803, USA}
\author{Ka-Ming~Tam}
\affiliation{Department of Physics and Astronomy, Louisiana State University, Baton Rouge, Louisiana 70803, USA}
\author{Mark Jarrell}
\affiliation{Department of Physics and Astronomy, Louisiana State University, Baton Rouge, Louisiana 70803, USA}
\author{Juana Moreno}
\affiliation{Department of Physics and Astronomy, Louisiana State University, Baton Rouge, Louisiana 70803, USA}
\author{Fakher Assaad}
\affiliation{Institute for Theoretical Physics and Astrophysics, University of W\"urzburg, Germany}
\author{Andy McMahan}
\affiliation{Lawrence Livermore National Laboratory, University of California, Livermore, California 94550, USA}   
\binoppenalty=10000

\begin{abstract}
This paper reports dynamical mean field calculations for the periodic Anderson model in which the conduction 
band is coupled to phonons.  Motivated in part by recent attention to the role of phonons in the 
$\gamma$-$\alpha$ transition in Ce, this model yields a rich and unexpected 
phase diagram which is of intrinsic interest.  Specifically, above a critical value 
of the electron-phonon interaction, a first order transition with two coexisting phases develops in the 
temperature-hybridization plane, which terminates at a second order critical point.  The coexisting phases display the 
familiar Kondo screened and local moment character, yet they also exhibit pronounced polaronic and bipolaronic 
properties, respectively.
\end{abstract}

\maketitle

The Periodic Anderson Model (PAM) and its impurity variant have played pivotal roles in elucidating 
the nature of Kondo screening as the techniques of many-body theory have improved~\cite{Hewson-1993, Jarrell-1993}. 
Perhaps its most noted application has been the Kondo Volume Collapse scenario for understanding the 
unique isostructural $\gamma$-$\alpha$ transition in Ce, with its very large 15\% volume 
change~\cite{Allen-1982, Allen-1992}.  The relative merits of this perspective versus the Mott transition 
scenario~\cite{Johansson-1974} are still under debate, although both focus on critical 4$f$-electron 
correlation effects, and the finite temperature predictions are rather similar~\cite{Held}. The 
PAM exhibits a smooth crossover from a local moment region with Curie Weiss susceptibility 
($\gamma$-like) to a region with Kondo screened 4$f$ moments and a paramagnetic susceptibility ($\alpha$-like), 
as a function of increasing hybridization between the 4$f$ and valence electrons~\cite{Huscroft}. Although 
the  PAM also predicts a first order transition given proper consideration of the Maxwell 
construction of the free energy versus volume curves~\cite{Allen-1982}, it requires modifications like some $f-f$
hybridization~\cite{Medici} so as to display a first order phase transition with two coexisting phases at the same 
hybridization in the temperature-hybridization plane.

Over the past decade attention in the Ce literature has shifted to an appreciation that a significant 
fraction of the total entropy change across the transition may be due to 
phonons~\cite{Manley-2003, Jeong, Decremps, Antonangeli, Johansson-2009, Lipp, Krisch-2011}. However, 
studies focusing on the effect of phonons on the PAM are very 
limited~\cite{Ono-2005, Hotta-2008, Nourafkan-2009, Raczkowsk-2010, Bodensiek-2010}. Prior studies
either are constrained to ground state calculation or do not explore possible phase transitions in detail. 
To this end, we are motivated here to consider the PAM with Holstein phonons~\cite{Holstein, Freericks, Capone, Koller}. 
Since the coupling of phonons to the $f$-electrons can lead to loss of local moments via electron condensation, 
we have chosen to couple the phonons to the conduction electrons in the present work. 
We find that the electron-phonon interaction above a critical 
strength induces a first order transition in the temperature-hybridization plane for the PAM-Holstein 
model. Strikingly the electron-phonon interaction also creates polaronic behavior in the 
Kondo screened phase and bipolaronic behavior in the local moment phase. This intriguing phase diagram is 
explored in the remainder of the present paper.

The Hamiltonian of the PAM-Holstein model is:
\vspace{-0.1in}
\begin{eqnarray}
H=H_{0}+H_{U}+H_{\rm{e-ph}},
\end{eqnarray}
\vspace{-0.2in}
\begin{eqnarray}
H_{0}=-t\sum_{\left\langle i,j\right\rangle,\sigma}\left(c_{i,\sigma}^{\dagger}c_{j,\sigma} +c_{j,\sigma}^{\dagger}c_{i,\sigma}\right)+\epsilon_{f} \sum_{i,\sigma}f_{i,\sigma}^{\dagger}f_{i,\sigma}  \nonumber \\ 
+V \sum_{i,\sigma}\left(c_{i,\sigma}^{\dagger}f_{i,\sigma}+f_{i,\sigma}^{\dagger}c_{i,\sigma}\right)+  \sum_{i}\left(\frac{P_{i}^{2}}{2m}+\frac{1}{2}kX_{i}^{2}\right) \nonumber
\end{eqnarray}
\vspace{-0.2in}
\begin{eqnarray}
H_{U}=U \sum_{i}n_{i,\uparrow}^{f}n_{i,\downarrow}^{f} \nonumber
\end{eqnarray}
\vspace{-0.2in}
\begin{eqnarray}
H_{\rm{e-ph}}=g \sum_{i,\sigma}n_{i,\sigma}^{c}X_{i}, \nonumber
\end{eqnarray}
where $c_{i,\sigma}$, $c^\dagger_{i,\sigma}$ ($f_{i,\sigma}$, $f^\dagger_{i,\sigma}$) are the creation and annihilation
operators of the conduction ($f$-level) at site $i$ and spin $\sigma$;
$n_{i,\sigma}^{c}=c^{\dagger}_{i,\sigma} c_{i,\sigma}$ and $n_{i,\sigma}^{f}=f_{i,\sigma}^{\dagger}f_{i,\sigma}$ 
represent the occupation of the $c$ and $f$-electrons, respectively; $t$ is
the nearest neighbor hopping; $\epsilon_f$, the on-site
energy of the $f$-level; $V$, the hybridization between conduction
and localized electrons; the on-site Hubbard interaction is $U$; $g$
is the electron-phonon coupling; $X_i$, the lattice displacement at site $i$, and $P_{i}$ its conjugate momentum. 

We use the Dynamical Mean Field Theory (DMFT) \cite{Georges} on a  hypercubic lattice 
in infinite dimensions with Gaussian density of states 
$D(\epsilon)=\displaystyle \frac{1}{\sqrt{\pi}W}e^{ -\left(\frac{\epsilon}{W}\right)^{2} }$.
 We set the bandwidth $W=1.0$ as the unit of energy. 
In Ce the Fermi energy is about $6000$ K and the Debye frequency is $110$-$160$K~\cite{Manley-2003,Krisch-2011}, 
therefore we set the phonon frequency $\omega_{0}=0.01$ at $1\%$ of bandwidth.  The total electronic density is fixed 
at $n=1.8$ by tuning the chemical potential at each iteration of the DMFT cycle. The Hubbard interaction is $U=4.0$, and we adjust 
$\epsilon_{f}$ so that $n_f=1$ at $T=0.1$ to ensure that a local moment is present at large temperatures. Therefore  
the  data we show are for $n_{f} \sim 1.0$ and $n_{c} \sim 0.8$.  The Continuous Time Quantum Monte Carlo (CTQMC)~\cite{Rubtsov}, 
generalized for electron-phonon coupling~\cite{Assaad}, is employed as the impurity solver.

\begin{figure}[h]
\centerline{
\includegraphics[height=0.35\textheight,width=0.3\textwidth, viewport= 80 10 600 780,clip,angle=270]{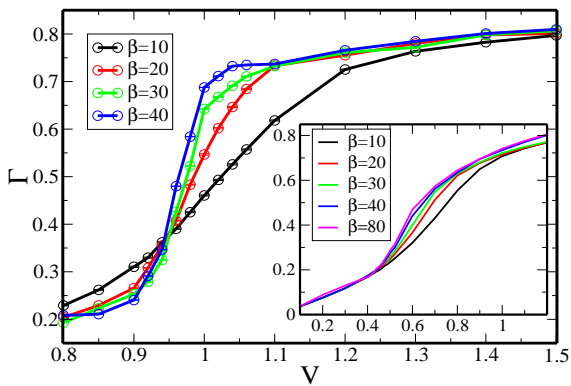}}
\vspace{-0.1in}
\caption{(color online) Isothermal scan of the hybridization factor $\Gamma=\left\langle c^{+}_0f_0+h.c.\right\rangle$  
as a function of $V$ for $g^{2}/2k=1.0$. $\Gamma$ increases monotonically with $V$. As the temperature decreases, 
$\Gamma$ vs.\ $V$ becomes steeper with a diverging slope near $V_c \sim 0.96$. Inset: The 
isothermal scan of the hybridization factor $\Gamma$ as a function of $V$ at $g^{2}/2k=0.49$. Notice that the critical behavior
has disappeared.}
\label{Hv_highT}
\vspace{-0.15in}
\end{figure}

Fig.~\ref{Hv_highT} displays the local hybridization factor $\Gamma=\left\langle c^{+}_0f_0+h.c.\right\rangle$  
(here 0 denotes the impurity site) as a function of $V$ for 
$g^{2}/2k=1.0$ and different values of inverse temperature, $\beta$.   As the temperature 
decreases, the slope of the $\Gamma$ vs.\ $V$ curve becomes progressively larger, which indicates 
the system is approaching a critical point. Interestingly,  the curves approximately cross for a critical hybridization 
of $V_{c} \sim 0.96$.  
The inset of Fig.~\ref{Hv_highT} shows $\Gamma$ vs.\ $V$ at $g^{2}/2k=0.49$.  
Notice that for this value of the coupling the slope does not become steeper  
as the temperature decreases, and the  crossing disappears. This indicates that the 
corresponding susceptibility reaches a plateau as a function of the temperature and the critical behavior 
is lost.  We believe that $g^{2}/2k=0.49$ is a lower bound 
for the critical value of the electron-phonon coupling.

\begin{figure}[bth]
\centerline{
\includegraphics[height=0.21\textheight,width=0.39\textwidth, viewport= 0 0 720 500,clip]{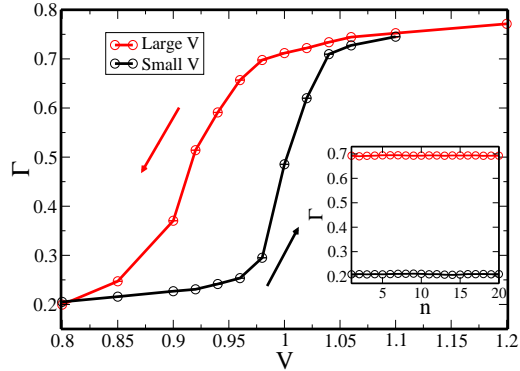}}
\vspace{-0.1in}
\caption{(color online) Hysteresis of $\Gamma$ vs.\ $V$ for $T=0.0167$, $g^{2}/2k=1.0$. The black line 
represents the small $V$ branch of the hysteresis for which the self-energy of the previous simulation is used 
to initiate the calculation for the next larger value of $V$. While the red line represents the large $V$ branch where
starting with $V=1.2$ we use the output of the previous simulation to initiate the computation at 
the next lower value of $V$.   Inset: $\Gamma$ as a function of the DMFT iteration number $n$ for $V=0.96$, $T=0.0167$. The black (red) 
symbols represent the small (large) $V$ branches.}
\label{Hysteresis}
\vspace{-0.15in}
\end{figure}

When the temperature is further decreased to $T=0.0167$ ($\beta=60$), $\Gamma$ vs.\ $V$ displays a hysteresis loop as shown in 
Fig.~\ref{Hysteresis}. The red line is obtained by starting at the large $V$ side ($V=1.2$), and using the output self-energy 
to initiate the simulation for the next smaller $V$. On the other hand, we obtain the black 
line by starting at $V=0.8$ and using the output self-energy as the input for the next larger value of $V$.  The 
coexistence of two solutions for the same value of $V$ at $T=0.0167$ is a direct evidence of a first order phase transition. 
The absence of such a hysteresis at higher temperatures indicates that the first order transition ends at a second order terminus ($V_{c},$ $T_{c}$). 

For the same parameters, $V=0.96$, $g^{2}/2k=1.0$, $\omega_{0}=0.01$, and $U=4.0$,
we also perform a series of isothermal scans on the chemical potential to study
the relationship between the total electron density $n=n_{c}+n_{f}$
and the chemical potential $\mu$. As long as the temperature is not
below $T=0.0167$, the compressibility $\displaystyle \frac{dn}{d\mu}$ shows no
tendency to diverge. This indicates the phase transition here is not
compressibility driven.

\begin{figure}[bth]
\centerline{
\includegraphics*[height=0.35\textheight,width=0.3\textwidth, viewport=80 0 600 780, clip,angle=270]{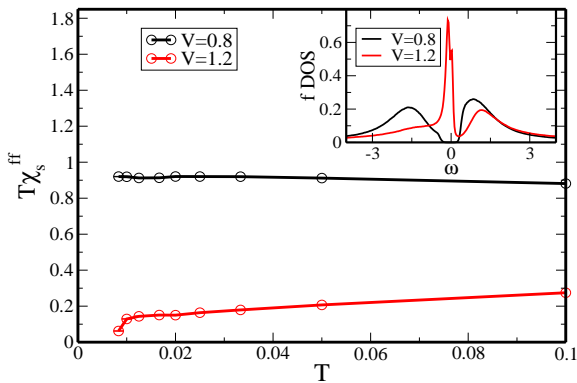}}
\vspace{-0.2in}
\caption{(color online) Temperature times the local $f$-orbital spin susceptibility, $T\cdot\chi_{s}^{ff}$, 
as a function of temperature for $g^{2}/2k=1.0$. 
For $V=0.8$ (black line), $T\cdot\chi_{s}^{ff}$ approaches a constant value as $T\rightarrow0$
indicating an unscreened moment.  For $V=1.2$ (red line), $T\cdot\chi_{s}^{ff}$ converges to zero indicating  the local moment
is screened. Inset: The $f$-electron DOS at $T=0.01$.  The Kondo peak found for
$V=1.2$ (red line), but absent for $V=0.8$ (black line) is consistent with the screening and unscreening scenarios in the main panel.}
\label{Chi_ff_s_T}
\vspace{-0.2in}
\end{figure}

In Fig.~\ref{Chi_ff_s_T} we show the temperature times the local $f$-orbital spin susceptibility, $T\cdot\chi_{s}^{ff}$,
versus temperature.  As $T$ approaches zero $T\cdot\chi_{s}^{ff}$ is roughly constant for $V=0.8$, while it goes to zero 
for $V=1.2$. This indicates that at $V=0.8$ the $f$-electrons display a robust local
moment and paramagnetic local susceptibility with $1/T$ dependence, while at $V=1.2$ the $f$ local moments are 
quenched.  The inset of Fig.~\ref{Chi_ff_s_T} shows the $f$-orbital density of states (DOS)
at $T=0.01$. Notice that at $V=0.8$ there is a gap across the Fermi level, while at $V=1.2$ a
Kondo resonance peak appears. The screening of the local moment in the large $V$ region  
is a consequence of the singlet formation between $c$ and $f$-electrons.

\begin{figure}[bth]
\centerline{
\includegraphics*[width=0.45\textwidth]{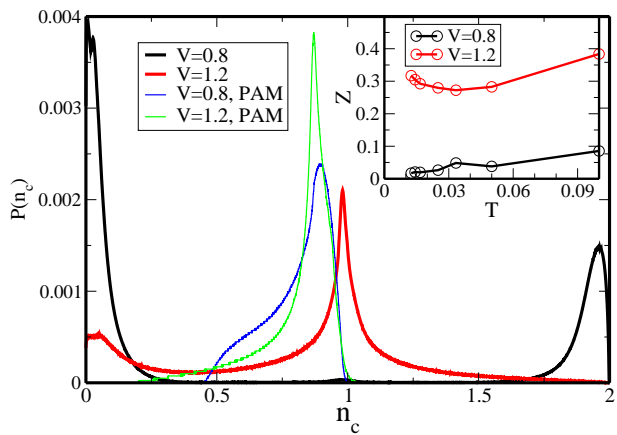}}
\vspace{-0.15in}
\caption{(color online) Occupancy distribution histogram of $c$-orbital $P(n_{c})$ for $V=0.8$ (black line) and $1.2$ 
(red line), $T=0.0167$ and $g^2/2k=1.0$. For comparison, $P(n_{c})$ of the PAM without electron-phonon coupling is
plotted as well: $V=0.8$ (blue line) and $V=1.2$ (green line). Inset:
the quasiparticle Z factor as a function of temperature for $V=0.8$ (black line) and $1.2$ (red line).}
\label{hist_Z}
\vspace{-0.2in}
\end{figure}

The main panel of Fig.~\ref{hist_Z} 
shows the occupancy distribution histogram of the $c$-electrons, $P(n_{c})$, at $T=0.0167$.
$P(n_{c})$ 
has been used to illustrate bipolaron formation~\cite{Assaad}. At $V=0.8$
the $c$-orbital electrons are in a bipolaronic state, which is characterized
by the oscillation between zero and double occupancy.
While for $V=1.2$, the $c$-electrons are in a polaronic state, where the
occupancy oscillates between zero and one. For the PAM, without electron-phonon coupling, 
the structure of $P(n_{c})$ is totally different.  Here there is only one peak at roughly
the $c$-electron filling $n_{c}=0.8$, and $P(n_{c})$ quickly decays to zero for $n_{c}$ 
away from this filling. In the inset, the quasi-particle fraction Z is plotted as a function 
of temperature. The quasi-particle fraction is calculated for the lower quasiparticle band 
at the Fermi level using a generalization of the single band formulation~\cite{Hess}.  The main component of this approach 
is to make the replacement $\displaystyle \frac{dRe\Sigma(\omega)}{d\omega} \vert_{\omega=0} \approx \frac{\mathrm{Im}\Sigma(i \pi T)}{\pi T}$, 
which becomes exact at zero temperature.  As $T\rightarrow0$, Z goes to zero for $V=0.8$ indicating non-Fermi liquid behavior, 
while it converges to a finite value for $V=1.2$, the signature of Fermi liquid formation.  

We find that this Kondo singlet to local moment phase transition 
remains for a large range of parameters, like adjusting the 
total filling to $n=1.6$, changing the Hubbard interaction to $U=3.8$ and increasing 
the phonon frequency to $\omega_{0}=0.02$ and $\omega_{0}=0.05$, while keeping $g^2/2k$ fixed.
For these different parameters, we find that the isothermal $\Gamma$ vs.\ $V$ curves
still cross and their slopes diverge at a critical value of the hybridization, $V_c$, as the temperature 
is decreased. We also find that $V_c$ changes roughly linearly with $g^2/2k$.

\begin{figure}[ht!]
\begin{center}{
\includegraphics[height=0.13\textheight,width=0.5\textwidth,clip]{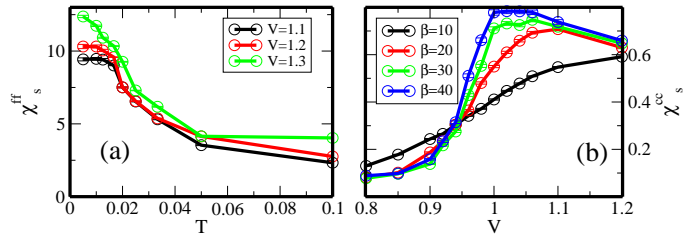}}
\vspace{-0.15in}
\caption{(color online) Panel (a) shows
the $f$-orbital time integrated local spin-spin correlation,
$\chi_{s}^{ff}$, as a function of temperature for $V=1.1$, $1.2$ and $1.3$.
Panel (b) shows the $c$-orbital time integrated local spin-spin correlation
function, $\chi_{s}^{cc}$, as a function of $V$ for different temperatures.}
\label{Chi_cc_and_ff}
\end{center}
\vspace{-0.2in}
\end{figure}

In Fig.~\ref{Chi_cc_and_ff}(a) the time integrated local $f$-orbital spin-spin correlation
function, $\chi_{s}^{ff}$, is plotted as a function of temperature for $V=1.1, 1.2$ and $1.3$. We identify
the Kondo scale $T_{K}$ as the energy where $\chi_{s}^{ff}$ falls to around half of its low temperature
value.  We find that $T_{K}$ changes very little as $V$ increases, so the line $V$ vs.\ $T_{K}$ 
should have a large slope. 
Fig.~\ref{Chi_cc_and_ff}(b) shows the 
time integrated local $c$-orbital spin-spin correlation function, $\chi_{s}^{cc}$ vs.\ $V$, at different temperatures,
where large values reflect the $c$-electron spin degeneracy in the polaronic state in contrast to the 
small susceptibility for the spinless bipolarons. For $V<0.96$ the  curves 
almost overlap for all $T<0.1$. In fact, the corresponding $c$-electron occupancy histograms (not shown)
show an obvious bipolaronic double peak feature even 
at relatively high temperatures like $T=0.1$. If we 
define $T^{*}$ as the energy where bipolaron formation begins, then the line $T^{*}$ vs.\ $V$ must be 
nearly horizontal.

We have also calculated the renormalized phonon frequency.  At $T=0.025$ it is 
roughly constant for hybridization $V > 0.96$; however, it drops precipitously for 
$V < 0.96$, decreasing by half when $V=0.8$.  This behavior 
softens with increasing temperature, e.g., a more gradual decrease begins for $V < 1.2$ at $T=0.1$. This indicates an
important temperature dependence of the phonons properties.  Indeed the analysis in~\cite{Lipp} for Ce found that the temperature 
dependence of the phonons was a critical factor for obtaining a significant phonon contribution to 
the entropy change across the $\gamma$-$\alpha$ transition~\cite{Manley-2003, Jeong, Decremps, Antonangeli, Johansson-2009, Lipp, Krisch-2011}.

\begin{figure}[t!]
\centerline{
\includegraphics*[height=0.27\textheight,width=0.36\textwidth,viewport=00 0 580 580,clip]{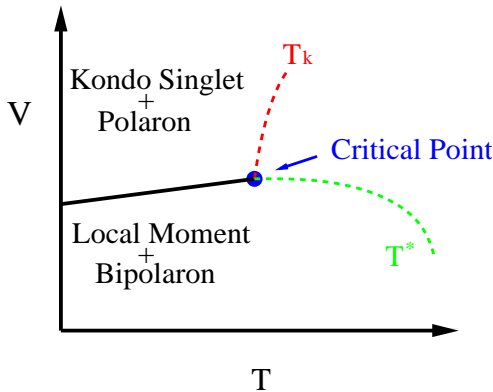}}
\vspace{-0.2in}
\caption{(color online) Schematic $V$ vs.\ $T$ phase diagram. The solid black line represents the first order phase transition which 
separate the local moment-bipolaron phase for small $V$ from the Kondo singlet-polaron phase for large $V$. This 
first order phase terminates at a second order critical point. The red dashed line coming out of the critical 
point represents the Kondo scale $T_{K}$ and the green dashed line the bipolaron energy scale $T^{*}$.}
\label{phase_diagram}
\vspace{-0.25in}
\end{figure}

Fig.~\ref{phase_diagram} is a schematic summary of our findings. Two phases, local moment-bipolaron and 
Kondo singlet-polaron, are separated by a first order transition line, which terminates at a second order 
critical point $(V_{c},T_{c})$. The positive slope of the $V$ vs.\ $T$ first order transition line is a consequence of 
a Clausius-Clapeyron-like relation where hybridization $V$ is the intensive analog of pressure. 
There is no broken symmetry between these two phases as we can move adiabatically 
from one to another by wandering around the critical point.  Both phases are destroyed by increasing the temperature. 
In order to have such a first order phase transition, the electron-phonon coupling on the $c$-band must be larger 
than a certain critical value. The fact that the critical temperature  is a function of electron-phonon coupling
implies that the critical point touches zero temperature at some $g_c$, where the first order phase transition 
becomes a quantum phase transition tuned by $V$.

In conclusion, when the conduction band of the periodic Anderson model is coupled to phonons, one 
obtains a rich and unexpected phase diagram.  Above a critical strength of the electron-phonon coupling
a first order transition  with two coexisting phases develops in the temperature-hybridization plane. This transition terminates 
at a second order critical point. These coexisting phases correspond to the familiar Kondo screened and 
local moment regions of the PAM, yet, they additionally exhibit pronounced polaronic and bipolaronic behavior, respectively.  
While the PAM and its impurity variant have been paradigms for the $\gamma$-$\alpha$ transition in Ce, additional electronic 
bands not considered here might be needed in a generalization of the present PAM-Holstein model to more completely explain 
the volume collapse. Nonetheless, the present simple model illustrates 
a fundamental principle relevant to the electron-phonon interaction in Ce. The Kondo temperature, which measures the critical 
energy scale of hybridization between  4$f$ and valence electrons, has a roughly exponential volume dependence leading to 
an order of magnitude increase from $\gamma$- to $\alpha$-Ce~\cite{Allen-1982, Allen-1992}.  This scale is comparable to that 
of the lattice vibrations (Debye temperature)~\cite{Krisch-2011} only in the $\gamma$ phase of Ce, and so 
it is no accident that the present work finds the most dramatic manifestations of the electron-phonon interaction, a 
bipolaronic state with significant phonon softening, in this local moment region.


We thank Thomas Pruschke for valuable discussions. This work is supported by 
NSF OISE-0952300 (PZ, PR, JM), DOE SciDAC grant DE-FC02-06ER25792 (KMT, MJ), and at LLNL by DOE 
contract DE-AC52-07NA27344 (AM). This work used the Extreme Science and Engineering 
Discovery Environment (XSEDE), which is supported by National Science Foundation 
grant number DMR100007.  
\vspace{-0.25in}

\end{document}